\documentclass[preprint,12pt]{elsarticle}
\biboptions{sort&compress}

\usepackage{graphicx,xcolor}
\usepackage{amsmath,amssymb}
\usepackage{hyperref}

\usepackage{epstopdf}
\usepackage{algorithm}
\usepackage[noend]{algpseudocode}
\algblockdefx{ParallelForEach}{EndParallelForEach}[1]{\algorithmicfor~\textbf{each} #1 \textbf{in parallel}}{\textbf{end parallel for}}
\algtext*{EndParallelForEach}

\journal{Computational Materials Science}

\begin{document}

\begin{frontmatter}

\title{Quantized bounding volume hierarchies for neighbor search in molecular simulations on graphics processing units}

\author[utaustin]{Michael P. Howard\corref{corrauth}}
\cortext[corrauth]{Corresponding author}
\ead{mphoward@utexas.edu}

\author[princeton]{Antonia Statt}
\author[princeton]{Felix Madutsa}
\author[utaustin]{Thomas M. Truskett}
\author[princeton]{Athanassios Z. Panagiotopoulos}

\address[utaustin]{McKetta Department of Chemical Engineering, University of Texas at Austin, Austin, Texas 78712}
\address[princeton]{Department of Chemical and Biological Engineering, Princeton University, Princeton, New Jersey 08544}

\begin{abstract}
We present an algorithm for neighbor search in molecular simulations on graphics processing units (GPUs) based
on bounding volume hierarchies (BVHs). The BVH is compressed into a low-precision, quantized representation to
increase the BVH traversal speed compared to a previous implementation.
We find that neighbor search using the quantized BVH is roughly two to four times
faster than current state-of-the-art methods using uniform grids (cell lists) for a suite of benchmarks for
common molecular simulation models. Based on the benchmark results, we recommend using the BVH instead of a
single cell list for neighbor list generation in molecular simulations on GPUs.
\end{abstract}


\end{frontmatter}

\section{Introduction}
Molecular simulation has become a valuable tool to provide insights on microscopic structures and processes,
to predict self-assembly and phase behavior, and to rapidly explore parametric design spaces.
The wide-spread use of molecular dynamics (MD) simulations \cite{Allen:1991,Frenkel:2002chF} in computational materials research
is due in part to the proliferation of optimized open-source MD software packages \cite{Plimpton:1995fc,Phillips:2005ts,Hess:2008tf,Anderson:2008vg}. The introduction of
massively-parallel computing architectures like the graphics processing unit (GPU) significantly expanded
the length and time scales accessible in MD simulations \cite{Anderson:2008vg,Friedrichs:2009,Colberg:2011de,Gotz:2012,Abraham:2015}, and most MD packages are now GPU-accelerated to some extent.

Particles in molecular simulations commonly interact through short-ranged, pairwise potentials. The prototypical
example is the Lennard-Jones potential,
\begin{equation}
U(r) = 4 \varepsilon \left[ \left(\frac{\sigma}{r} \right)^{12} - \left(\frac{\sigma}{r} \right)^6 \right], \label{eq:lj}
\end{equation}
where $r$ is the distance between particles, $\varepsilon$ is the interaction energy, and $\sigma$ is the
particle diameter. For computational efficiency in MD simulations, eq.~\ref{eq:lj} is usually truncated and shifted to zero at a
distance $r_{\rm c}$ that is long enough that the properties of interest are not significantly affected by the neglected attractions.
A disordered collection of particles interacting only by eq.~\ref{eq:lj} is called a Lennard-Jones fluid.

\begin{figure}
    \centering
    \includegraphics{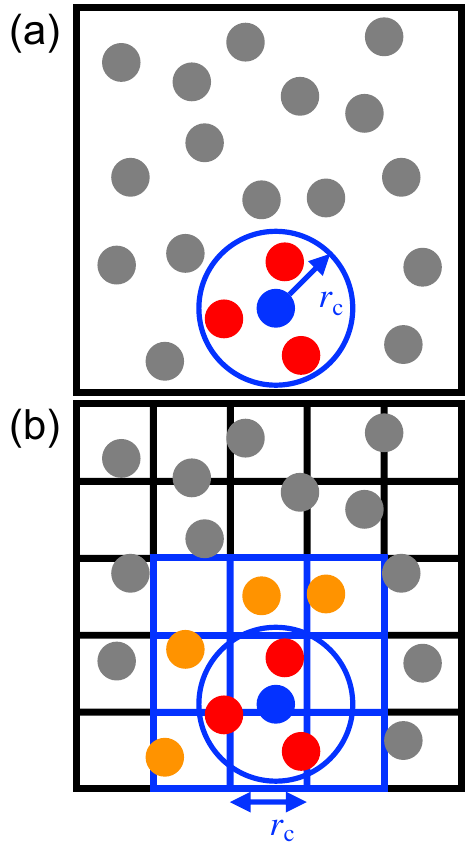}
    \caption{(a) Illustration of neighbor search. Red particles are neighbors of the blue particle within
             distance $r_{\rm c}$, while the gray particles are not neighbors. (b) Grid algorithm for neighbor search using cell width $r_{\rm c}$.
             The blue particle checks for neighbors from only the blue cells. The red particles are tested
             and found to be neighbors, the orange particles are tested but are not neighbors,
             and the gray particles are skipped.}
    \label{fig:neigh}
\end{figure}

The calculation of pairwise interactions is usually the most computationally demanding part
of a molecular simulation. The complexity of exhaustively evaluating all pairs from $N$ particles is $O(N^2)$;
however, many of these interactions are trivially zero when $r > r_{\rm c}$.
The challenge is then to efficiently determine the subset of particles within a distance $r_{\rm c}$
of a given particle (Fig.~\ref{fig:neigh}a). The pair interactions can be computed directly with these particles,
or the subset can be saved to construct a Verlet (neighbor) list \cite{Frenkel:2002chF};
both problems are fundamentally a \textit{neighbor search}.

Neighbor search has historically been performed in molecular simulations using a grid
(``cell list'') approach \cite{Frenkel:2002chF}. The grid subdivides space uniformly into cells,
and particles are binned into the cells (Fig.~\ref{fig:neigh}b). A pairwise
distance check between particles is only done for a small number of adjacent cells, reducing the search complexity
to $O(Nm)$ with $m$ being the average number of particles in a cell (usually $m \ll N$). This algorithm
works well on CPUs \cite{Plimpton:1995fc,intVeld:2008hx} and has also been successfully adapted to GPUs \cite{Colberg:2011de,Glaser:2015cu,Rushaidat:2015}.

Despite the simplified complexity of the grid search, many distance checks are still wasted by
this scheme \cite{intVeld:2008hx,Howard:2016wq}. If the cell size is equal to $r_{\rm c}$, the volume of the neighbor-search sphere is only 16\% of
the explored cell volume in three dimensions. Decreasing the cell size (increasing the number of cells) to reduce this waste
results in significant overhead on GPUs \cite{Howard:2016wq}. Some alternatives \cite{Friedrichs:2009,Gotz:2012,Pall:2013,Abraham:2015}
to standard cell lists trade more distance checks for reduced overhead in an attempt to fully leverage the parallelism of the GPU.

In computer gaming and ray tracing, tree structures like bounding volume hierarchies (BVHs), $k$-d trees,
and octrees are often used instead of a uniform grid to efficiently detect collisions and intersections \cite{ericson2004real}.
These algorithms group particles that are nearby in space into a tree-like hierarchy that can be searched
(traversed) by simple intersection tests. Such trees have been successfully employed for neighbor search in
molecular simulations on CPUs \cite{Grudinin:2010jh,Artemova:2011cx,Anderson:2016,Tortora:2017,Chen:2017}.

Recently, some of us proposed using a BVH to perform neighbor search in molecular simulations on GPUs \cite{Howard:2016wq}. The BVH significantly
outperformed the grid for mixtures having interaction-range disparity, e.g., large colloidal particles dispersed
in a solvent, due to its lower traversal overhead for such mixtures. Unsatisfactorily, though, the BVH was often slower than the grid when the size disparity was smaller.
The BVH performed worst compared to the grid for a single-component Lennard-Jones fluid; the grid was nearly twice as
fast \cite{Howard:2016wq}. At the time, we recommended using the BVH for systems having
significant interaction range disparity, but the grid approach otherwise.

We have since revisited the BVH neighbor-search algorithm and made substantial improvements
to the BVH traversal. The key development is to employ a low-precision, quantized representation that compresses the internal BVH data
and reduces the traversal overhead. We performed a comprehensive set of benchmarks for single-component
fluids (the previous worst-case scenario) with the new algorithm using recent NVIDIA GPUs. We find that the
new BVH algorithm is comparable or superior to the grid method for nearly all benchmark configurations tested.
On current GPUs, neighbor search with the BVH is typically two to four times faster than with an equivalent
grid algorithm.

The rest of this article is organized as follows. We first provide an overview of the BVH neighbor-search algorithm
with a focus on molecular simulations and then describe the improvements we have made to our previous implementation
by compressing the BVH (Section ~\ref{sec:alg}). We then give details of the algorithm implementation
(Section \ref{sec:impl}) and test its performance for a suite of benchmarks based on
representative molecular simulation models for fluids (Section \ref{sec:perf}). Finally, we summarize our findings and suggest
avenues for further investigation (Section \ref{sec:conc}).

\section{Algorithm} \label{sec:alg}
\subsection{Bounding volume hierarchy}
\begin{figure}
    \centering
    \includegraphics{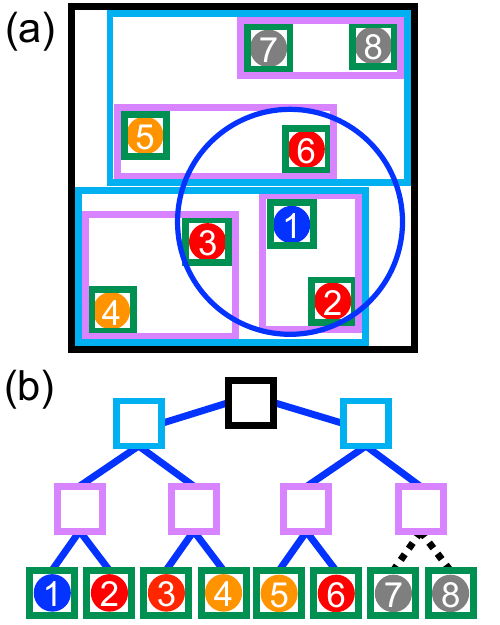}
    \caption{Schematic bounding volume hierarchy corresponding to configuration of particles inside blue cells of Fig.~\ref{fig:neigh}b.
        (a) Nearby particles (circles) are grouped into successively larger axis-aligned bounding boxes. The leaf nodes are
        green, the internal nodes are purple and blue, and the root node is black.
        (b) Hierarchical view of the bounding volumes in (a).
        Blue solid edges indicate branches that are visited
        during neighbor search for particle 1 using the shown large circle, while the black dotted edges are skipped.
        The leaf nodes containing the red particles are tested and found to be neighbors, the leaf nodes containing the orange
        particles are tested but are not neighbors, and the leaf nodes containing the gray particles are skipped. Note that although
        the leaf node for particle 1 is trivially visited during traversal, a particle is usually not considered a neighbor of itself.}
    \label{fig:bvh}
\end{figure}

A BVH is a tree structure that partitions a system by objects rather than space \cite{ericson2004real}. A schematic is shown
in Fig.~\ref{fig:bvh}a for a subset of the particles in Fig.~\ref{fig:neigh}. Each object (circles) is placed in a ``leaf''
node, and leaf nodes are successively merged into internal nodes, ending at the ``root'' node at the top of the tree. Each node has a bounding volume that encloses
all of its descendant objects. In this discussion, we assume that the tree is binary and each
internal node has exactly two children, but wider BVHs have also been used for ray-tracing applications \cite{Dammertz:2008,Ernst:2008,Ylitie:2017}.

Objects within a given volume can be found by traversing the BVH using a binary search \cite{ericson2004real}, illustrated in
Fig.~\ref{fig:bvh}b. To find neighbors of a given particle, a search volume is tested for overlap with the bounding volume of a node starting from
the root, which has the largest bounding volume in the tree and no ancestors.
Traversal proceeds to the children of that node if an overlap is detected; otherwise, that branch of the tree can be skipped.
Eventually, the traversal reaches a leaf node, where additional (or more expensive) calculations between the search
volume and the object in the leaf can be performed if needed.

In order to obtain good traversal performance, objects should be equally balanced between branches of the BVH, and objects that are spatially
local should also be grouped nearby in the hierarchy \cite{ericson2004real}. Typically, there is a tradeoff between the BVH
build time and the quality of the BVH for traversal, with the optimal balance depending on the number of traversals performed per build \cite{Karras:2013wb}.
In our previous work, we constructed a linear bounding volume hierarchy (LBVH) \cite{Lauterbach:2009js} using Karras's algorithm \cite{Karras:2012ur} that builds
the BVH from the bottom up to maximize parallelism on the GPU. Higher quality BVHs can be constructed by
alternative algorithms \cite{MacDonald:1990vk,Stich:2009,Garanzha:2011wv} or the LBVH can be refined with additional processing \cite{Karras:2013wb,Domingues:2015}, but both carry a significant
cost. We have found that the quality of the LBVH is sufficient to obtain good performance in our benchmarks (Section \ref{sec:perf}),
and so we have not attempted to implement these alternative BVH construction schemes.

We will briefly summarize the algorithm for constructing the LBVH and traversing it on the GPU for a molecular simulation;
additional details and pseudocode are available elsewhere \cite{Karras:2012ur,Howard:2016wq}. To build the LBVH, the simulation box is first subdivided into $2^{10}-1$ bins along each
Cartesian axis, with each bin coordinate represented by a 10-bit integer. The particles are assigned into a bin, given a
30-bit Morton code by interleaving the bin coordinates bitwise, and then sorted along a Z-order curve \cite{Morton:1966vb}. The
Morton codes are subsequently processed to construct a tree hierarchy \cite{Karras:2012ur},
and bounding volumes are fit for each node by walking up the tree from the leaf nodes. We enclose the nodes
in axis-aligned bounding boxes (AABBs) because they are simple to construct, have a small memory footprint, and
are easy to test for overlaps. The constructed BVH has $N-1$ internal nodes for $N$ objects (stored in
$N$ leaf nodes), and each internal node has exactly two children \cite{Karras:2012ur}.

We employ a stackless scheme to traverse the BVH (lines 10-19 of Algorithm~\ref{alg:bvh}) \cite{MacDonald:1990vk,Smits:1998ui,Torres:2009vr,Howard:2016wq}.
Every particle searches the BVH using a bounding volume centered around
the particle; we previously used an AABB with inscribed radius $r_{\rm c}$ as the search volume \cite{Howard:2016wq}. In order to
accommodate periodic boundary conditions in molecular simulations, the search volume is translated by appropriate combinations
of the lattice vectors, giving 27 total search volumes per particle for a three-dimensional periodic simulation box. When the search volume
overlaps an internal node, the traversal descends to the left child of the node. Otherwise, traversal advances along a
``rope'' to the next node in the tree to test. Traversal terminates when the rope advances past the last node
in the tree. We found in preliminary benchmarks that this stackless approach performed favorably compared to using
an explicitly-managed stack \cite{Aila:2009,Aila:2012}.

\subsection{Compressed BVH}
The total neighbor search time in our previous implementation was dominated by the BVH traversal. Moreover,
a significant fraction of the BVH build time was simply due to sorting the Morton codes. We accordingly focused
our efforts on improving the traversal. In particular, inspired by developments in ray tracing \cite{Mahovsky:2006,Kim:2010,Segovia:2010,Keely:2014,Vaidyanathan:2016,Ylitie:2017},
we considered ways to compress the BVH to reduce the data loaded per node, which is often a
bottleneck for GPUs.

\begin{figure}
    \centering
    \includegraphics{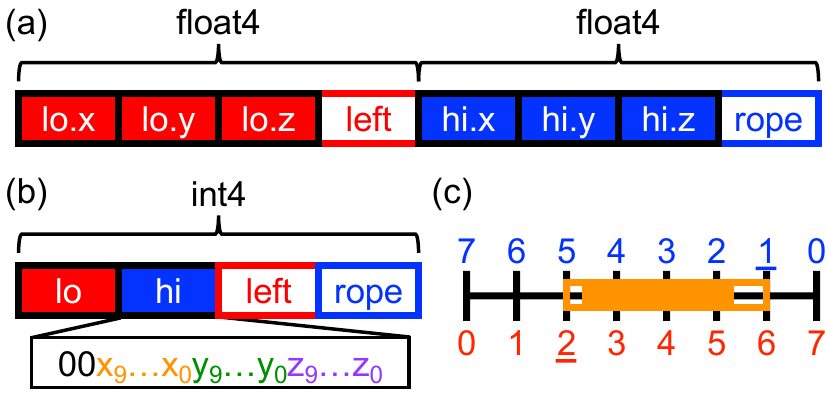}
    \caption{(a) BVH node layout similar to ref.~\citenum{Howard:2016wq}. lo and hi are lower and upper bounds of the AABB, while left
        and rope are indexes of nodes for traversal. (b) Compressed node layout for quantized bounds that requires only
        half the memory of the original layout (a). The box shows how the quantized bounds are concatenated into one 32-bit
        integer. (c) Schematic illustration of quantized lower (red) and upper (blue)
        bounds in $x$ for an AABB (filled orange).
        The underlined indexes give the integer bounds of the quantized AABB (open orange) on the
        discretized grid (black). float4 and int4 are CUDA data types that can be used to store the data, with each being 16 bytes \cite{nvidia:cuda}.}
    \label{fig:compress}
\end{figure}

The original memory layout of our BVH nodes \cite{Howard:2016wq} is shown in Fig.~\ref{fig:compress}a with slight modification. An AABB is defined by a lower and upper
bound in three dimensions, represented as floating-point values. Each node additionally stores the integer index of
its left child in the tree and its skip rope for traversal. When the node is a leaf, it has no left child, and the node instead
stores the index of the object in the leaf. In practice, this distinction is made by representing the indexes as 32-bit signed
integers, with nodes being positive and objects being negative.

Many MD simulations are conducted using 64-bit (``double'') floating-point precision for the particle data and
in key steps to ensure faithful integration of the equations of motion \cite{Colberg:2011de}. However, such high precision would be
wasteful for the AABB since only simple overlap tests are needed. For smaller memory demands and better performance,
the AABB bounds can instead be stored in lower precision, e.g., 32-bit (``single'') floating-point precision.
A node with a single-precision AABB is then only 32 bytes (Fig.~\ref{fig:compress}a) instead of the minimum of
56 bytes that would be needed for a node with a double-precision AABB.

For correctness, the lower-precision AABB bounds must fully enclose the volume of the higher-precision AABB or object \cite{Vaidyanathan:2016}.
For example, to fit a single-precision AABB to a point particle represented in double precision, the lower bound of the
AABB is computed by rounding down the particle coordinates to the nearest representable single-precision value, while the upper bound
is computed by rounding up. The AABB then represents the particle position with uncertainty due to the loss of
precision. Any search volume overlapping the point particle must also overlap the AABB, and so correctness of the search
is ensured (no false negatives). However, there may be some overlaps with the AABB that do not correspond to actual overlaps
with the particle (false positives).

We propose a scheme to further compress the node memory beyond single-precision AABB bounds through appropriate
discretization of space into bins. The AABB bounds can then be stored in a compact, quantized integer representation using the bin coordinates (Fig.~\ref{fig:compress}c). This strategy is similar to previous
ones \cite{Mahovsky:2006,Keely:2014,Vaidyanathan:2016,Ylitie:2017} but uses a global coordinate system based on the root-node AABB that is better suited to binary trees and stackless traversal.
We subdivide the root-node AABB into $2^{10}-1$ bins along each dimension.
The lower and upper bounds of the AABBs in the BVH are snapped onto this grid, ensuring that lower bounds are rounded down and
upper bounds are rounded up for correctness \cite{Vaidyanathan:2016,Ylitie:2017}. These 10-bit integer representations can then be concatenated
into a 32-bit integer representing each bound (2 bits are unused). The entire node needs only 16 bytes (Fig.~\ref{fig:compress}b),
half the size of the node with a single-precision AABB.

More false-positive overlaps are expected for the quantized AABB than for the single-precision or double-precision AABB,
particularly when the discretization becomes large relative to $r_{\rm c}$. The number of false overlaps
affects traversal performance, and so there is a tradeoff between reducing the size of the node in this way and the BVH quality
for traversal. For typical simulation boxes of size $50\,\sigma$, the global discretization is roughly $0.05\,\sigma$,
leading to loss of precision in the first decimal place. We will characterize the effectiveness of this
discretization later in our benchmarks.

\subsection{Additional improvements}
In order to fully capitalize on the compressed BVH representation, we have further modified our original
BVH traversal algorithm \cite{Howard:2016wq}, presented in its new form in Algorithm~\ref{alg:bvh},
so that no additional distance checks are performed between particles once a leaf node is
reached. Instead, we simply accept particles as neighbors if the search volume overlaps the leaf in the traversal (line 14).
With this approach, the particle coordinates, which are typically stored in higher precision than the AABB,
no longer need to be loaded. However, a small number of false-positive neighbors will be identified,
which can be filtered or rejected in a later stage if desired.

\begin{algorithm}
    \caption{BVH neighbor search in a molecular simulation.} \label{alg:bvh}
    \begin{algorithmic}[1]
        \State $\{v_j\} \gets$ periodic image translation vectors
        \ParallelForEach{particle $i$}
        \State $x_i \gets$ position of particle $i$
        \State $b \gets$ integer with bit $j$ set to 1 if $x_i + v_j$ overlaps the root node
        
        \While{$b \ne 0$}
        \State $j \gets$ index of next set bit in $b$
        \State Set bit $j$ in $b$ to 0.
        \State $S \gets$ \Call{sphere}{$x_i+v_j$,$r_{\rm c}$}
        \State $n \gets$ root node
        
        \While{untested node $n$ remains}
        \State $A \gets$ \Call{aabb}{$n$}
        \If{$S$ overlaps $A$}
        \If{$n$ is a leaf}
        \State Process object in $n$ as a neighbor.
        \State $n \gets$ next node to test from rope
        \Else
        \State $n \gets$ left child of $n$
        \EndIf
        \Else
        \State $n \gets$ next node to test from rope
        \EndIf
        \EndWhile
        \EndWhile
        \EndParallelForEach
    \end{algorithmic}
\end{algorithm}

In order to limit the number of false-positive overlaps, we search the BVH using a spherical volume with radius $r_{\rm c}$
centered around each particle rather than the AABB with inscribed radius $r_{\rm c}$ from our previous work \cite{Howard:2016wq} (line 8). Using a sphere instead of an AABB decreases the search volume by roughly 50\%,
which reduces the number of false positives from, e.g., overlaps between corners of AABBs.
To perform the sphere--AABB overlap tests with internal and leaf nodes, the quantized AABB is first decompressed into a
single-precision floating-point representation (line 11). Rounding ensures that the node bounds are correctly reconstructed.
For best arithmetic instruction performance, the sphere is also represented in single precision regardless of the
precision of the particle data; the sphere radius is padded to account for uncertainty from
loss of precision when the particle data is double precision.

Overlap is then tested by finding the point $\mathbf{y}$ in the AABB nearest to the center of the
sphere $\mathbf{x}$ (line 12). If $\mathbf{a}$ and $\mathbf{b}$ are the lower and upper bounds of the AABB,
then the nearest point to $\mathbf{x}$ in the AABB is \cite{ericson2004real}
\begin{equation}
\mathbf{y} = \min(\max(\mathbf{x},\mathbf{a}), \mathbf{b}),
\end{equation}
where the minimum and maximum operations are applied component-wise. The sphere and the AABB overlap if
the distance from the center of the sphere to this point is less than the sphere radius, i.e., 
$|\mathbf{y} - \mathbf{x}| \le r_{\rm c}$.

The search sphere is translated by appropriate combinations of the simulation box lattice vectors to account for
periodic boundary conditions (line 8). We previously implemented this as a loop over images \cite{Howard:2016wq}, but this scheme may lead to
execution divergence on the GPU when some images do not overlap the BVH. As a minor optimization, we now test all images for overlap
with the root node and construct a bitset indicating which images overlap before beginning traversal (line 4). The next
image traversed for a particle can be selected from this bitset (line 6), skipping over nonoverlapping images, without wasting
iterations of the loop.

Finally, our previous implementation of the BVH algorithm lumped multiple particles into leaf nodes using the Z-order curve before the BVH was constructed \cite{Howard:2016wq}.
This procedure reduced the depth of the BVH for traversal, and we found that grouping 4 consecutive particles from the
Z-order curve per leaf was optimal in that implementation. However, subsequent analysis of the BVH structure revealed that
this procedure produced rare large AABBs when a grouping of 4 spanned a jump in the Z-order curve. The AABBs of internal nodes
enclosing such leaf nodes were correspondingly large, slowing traversal. A better approach would first construct the tree with one
particle per leaf and then collapse subtrees according to an appropriate heuristic \cite{Karras:2013wb}. In practice, we have found in our
benchmarks that the BVH performance is already good without any subtree collapse and leave this as a possible future
optimization.

\section{Implementation} \label{sec:impl}
We implemented the quantized BVH algorithm in a neighbor-search library \cite{neighbor} that uses HOOMD-blue (version 2.4.1) \cite{Anderson:2008vg,Glaser:2015cu} as a dependency.
HOOMD-blue is an open-source molecular dynamics package optimized for NVIDIA GPUs. The remainder of our discussion will accordingly use
terminology from the CUDA programming model \cite{nvidia:cuda}. HOOMD-blue can be configured to use either single- or double-precision
floating-point values to represent particles, and our implementation supports both data types.

We constructed an LBVH using Karras's
algorithm \cite{Karras:2012ur} with the CUB library \cite{cubweb} for sorting. Neighbor search was performed using Algorithm~\ref{alg:bvh} with the particles
in Z-order so that threads within the same CUDA warp traversed similar parts of the tree \cite{Howard:2016wq}. Arithmetic operations
for compressing and decompressing the AABBs and detecting overlaps were performed using IEEE 754-2008 round-down and round-up modes \cite{ieee754}
through the appropriate CUDA intrinsics \cite{nvidia:cuda}.

We compared the neighbor-search performance of the BVH to an equivalent uniform grid algorithm. To construct the grid, particles were binned into cells
of size $r_{\rm c}$ \cite{Howard:2016wq}, and the particles in each cell were identified by sorting \cite{Colberg:2011de}. This approach differs from
the current implementation in HOOMD-blue \cite{Glaser:2015cu}, where particles are inserted into cells using limited atomic operations. The
sorting approach produces a more memory-compact list of particles in cells than the atomic operations, and the particle list is additionally
deterministic if the sorting algorithm is stable. Both are desirable for molecular simulations and are also properties shared by the LBVH algorithm.
We found in selected benchmarks that the sorting approach with compact memory was roughly two times slower than HOOMD-blue's aggressively optimized atomic
approach; this appears to partially be an inherent cost of making an equivalent grid algorithm that is both deterministic and minimally memory consuming like the LBVH.

Grid traversal \cite{Glaser:2015cu,Howard:2016wq,Howard:2018} was implemented using $n$ threads per search
volume, where $n$ is a power of 2 smaller than the CUDA warp size. The $n$ threads cooperatively
process particles from adjacent cells with a stencil. They communicate using CUDA shuffle instructions \cite{nvidia:cuda} that allow threads within a
warp to efficiently read each others' registers. Particle data was sorted into the same order
as the grid to improve traversal speeds, and neighbor search was performed in this sorted order.
All distance evaluations were performed using the precision of the particle data (single or double precision).

\section{Performance} \label{sec:perf}
We tested the performance of both algorithms using disordered (fluid) configurations generated from MD simulations of commonly used models
(see below). We emphasize that these benchmarks gauge the performance of the neighbor search algorithm itself, which
is part of the total molecular simulation time along with, e.g., pair force evaluation and integration of the equations
of motion. We collected 10 independent configurations for each benchmark from constant-temperature,
constant-volume MD simulations in HOOMD-blue. The simulation time step was $0.005\,\tau$, where $\tau = \sqrt{m \sigma^2/\varepsilon}$
is the unit of time for particles of mass $m$. The temperature $T$ was controlled using a Langevin thermostat with friction
coefficient $0.1\,m/\tau$ \cite{Phillips:2011td}. The simulation box was cubic and periodic in all three dimensions, and its size $L$ was set
to obtain the desired density $\rho = N/L^3$ for $N$ particles.

We subsequently determined the average time to build and traverse a BVH or grid using these configurations. The traversal
counted the number of particles within a distance $r_{\rm c}$ of each of the $N$ particles in the configuration, subject
to the periodic boundary conditions. We first
ran 200 builds and traversals to allow runtime autotuners \cite{Glaser:2015cu} to determine the optimal CUDA kernel launch parameters.
We then disabled the autotuners and measured the average build and traversal times over 500 iterations. We repeated
this measurement 5 times, determined the median value, and averaged it for all configurations.
We ran all benchmarks using both single- and double-precision particle data representations.

The benchmarks were performed on recent NVIDIA Tesla GPUs commonly found in supercomputing centers:
K80, P100 for PCIe (16 GB), and V100 for PCIe (32 GB). We additionally tested performance on GeForce GTX 1080,
which is designed for computer gaming but is also often used for computations. Each GPU architecture has unique
features that are beyond the scope of this article to describe \cite{nvidia:cuda}.
Benchmarks programs for K80, P100, and V100 were compiled using CUDA 9.0 and GCC 6.4.0,
while the GTX 1080 benchmark used CUDA 8.0 and GCC 4.8.4. The benchmarks were run on a single GPU driven
by a single process; the algorithms presented here can be straightforwardly generalized to execute on multiple
GPUs through a domain decomposition strategy \cite{Glaser:2015cu}.

\subsection{Lennard-Jones fluid}
We first tested the performance of the algorithms for the Lennard-Jones fluid, which presented the most significant
challenges for our original BVH neighbor-search implementation \cite{Howard:2016wq}. We ran benchmarks with $r_{\rm c} = 3.0\,\sigma$
at both low ($\rho = 0.2/\sigma^3$) and high ($\rho = 0.8/\sigma^3$) densities to mimic a range of typical
simulation conditions. (The triple-point density of the Lennard-Jones fluid
is roughly $0.86/\sigma^3$ \cite{Ahmed:2009}.) Configurations with $N = 128,000$ particles were sampled every $10^4\,\tau$
at temperature $T=1.5\,\varepsilon/k_{\rm B}$ (above the critical temperature \cite{Potoff:1998,Panagiotopoulos:1994}),
where $k_{\rm B}$ is Boltzmann's constant. This number of particles
was sufficient to fully saturate the GPU with work in our simulations and benchmarks.

\begin{figure}[!htbp]
    \centering
    \includegraphics{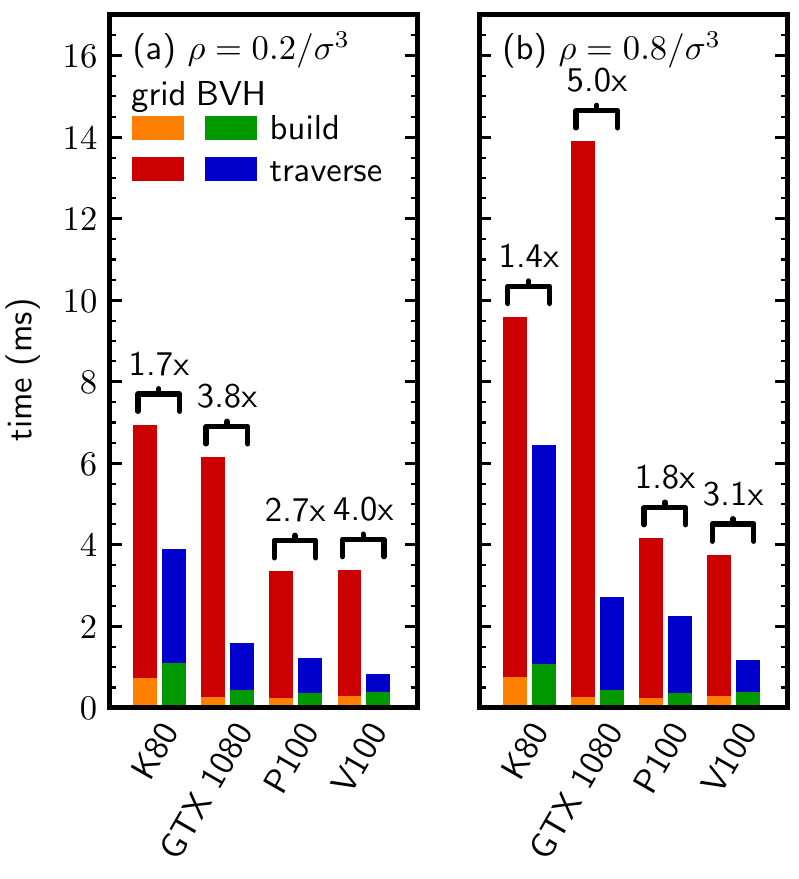}
    \caption{Lennard-Jones fluid neighbor-search benchmark for $N = 128,000$ particles represented in double precision at
        (a) $\rho = 0.2/\sigma^3$ and (b) $\rho = 0.8/\sigma^3$ with $T = 1.5\,\varepsilon/k_{\rm B}$.
        The traversal counted the number of neighbors within a cutoff $r_{\rm c} = 3.0\,\sigma$.
        Speedups are the ratios of the total times rounded down to the nearest tenth.}
    \label{fig:ljdouble}
\end{figure}

When the particle data was represented in double-precision, the BVH neighbor search was
consistently faster than the grid for all GPUs tested (Fig.~\ref{fig:ljdouble}). The time required to build the
LBVH was nearly twice the time to construct the grid, which can be attributed to the additional overhead of
building the tree hierarchy and fitting the bounding volumes after sorting the particles.
However, the traversal of the LBVH was much faster than for the grid, resulting in a significant net
speedup of the neighbor search. The speedup on the Tesla GPUs was larger at the lower density, for which the overhead
of accessing a cell from the grid is higher (cell occupancy is lower). However, the largest speedup was obtained
for the GTX 1080 at $\rho = 0.8/\sigma^3$ due to the low double-precision arithmetic instruction throughput of that GPU \cite{nvidia:cuda},
which limited the grid algorithm performance. In such cases, using the quantized BVHs with single-precision
arithmetic is highly advantageous.

\begin{figure}[!htbp]
    \centering
    \includegraphics{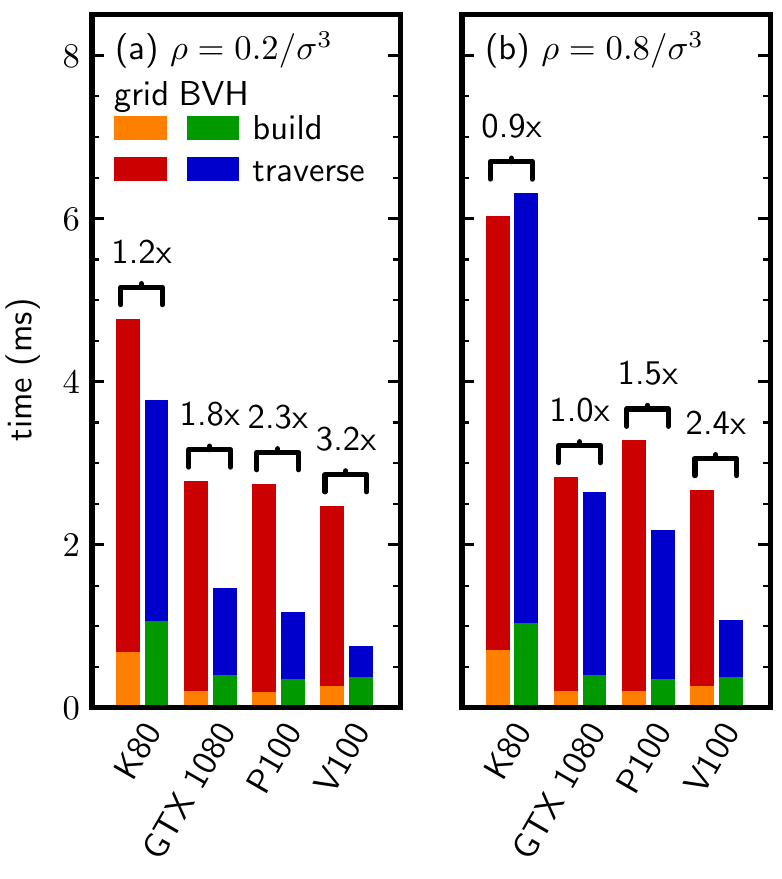}
    \caption{Lennard-Jones fluid neighbor-search benchmark for $N = 128,000$ particles represented in single precision at
        (a) $\rho = 0.2/\sigma^3$ and (b) $\rho = 0.8/\sigma^3$ with $T = 1.5\,\varepsilon/k_{\rm B}$.
        The traversal counted the number of neighbors within a cutoff $r_{\rm c} = 3.0\,\sigma$.
        Speedups are the ratios of the total times rounded down to the nearest tenth.}
    \label{fig:ljsingle}
\end{figure}

We repeated the same benchmark using single-precision particle data. We expected the grid performance to be more
sensitive to the precision of the particle data than the BVH because the grid does not use a low-precision
internal representation. Consistent with this expectation,  the BVH neighbor-search
performance was again comparable or superior to the grid (Fig.~\ref{fig:ljsingle}), but the speedups were generally
smaller than for the double-precision data (Fig.~\ref{fig:ljdouble}). This is likely due to two reasons: (1) the
single-precision arithmetic throughput of the GPU is higher than the double-precision throughput, resulting
in faster grid traversal, and (2) the quantized BVH node is no longer smaller than the single-precision particle
data; both are 16 bytes in our implementation. The first effect is most pronounced for the GTX 1080 when
comparing Figs.~\ref{fig:ljdouble} and \ref{fig:ljsingle}, while the second affects all GPUs tested. However, we
emphasize that on recent GPUs (P100, V100), the BVH neighbor search is still faster than the grid even for
single-precision particle data.

The previous benchmarks were run with $N$ sufficiently large that the GPU was fully saturated with work.
In smaller simulations or to obtain efficient strong scaling in multi-GPU simulations \cite{Glaser:2015cu}, it is important to have
good performance even when the number of particles becomes small. For the grid, this is aided by assigning
$n$ CUDA threads to cooperatively process the neighbors of each particle \cite{Glaser:2015cu}. The optimal value of $n$
depends on $N$, $\rho$, and the GPU, and can be determined at run time. On the other hand, the BVH traversal
(Algorithm~\ref{alg:bvh}) uses only one thread per particle regardless of $N$ and $\rho$, which could
fail to saturate the GPU with work at small $N$.

\begin{figure}[!htbp]
    \centering
    \includegraphics{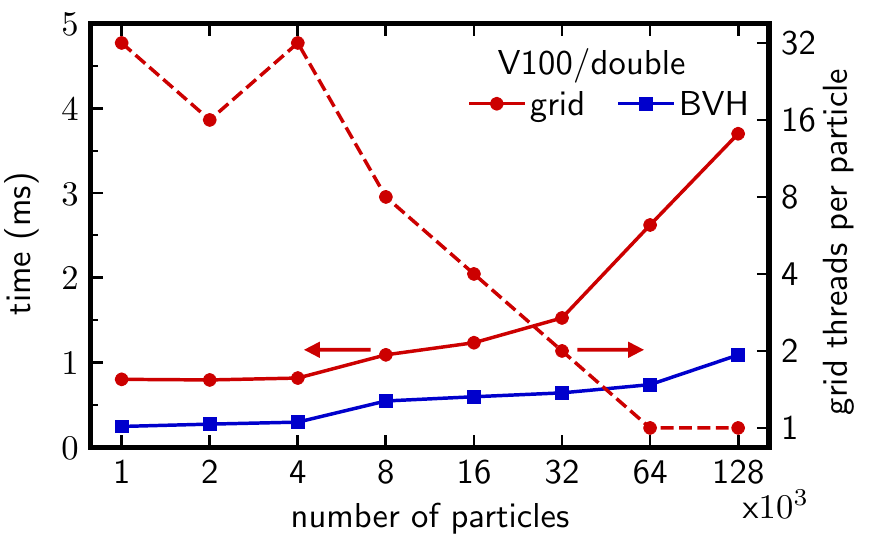}
    \caption{Lennard-Jones fluid benchmark on Tesla V100 for PCIe (32 GB) for $N$ particles represented in double precision
             at density $\rho = 0.6/\sigma^3$ and $T = 1.5\,\varepsilon/k_{\rm B}$. The traversal
             counted the number of neighbors within a cutoff $r_{\rm c} = 3.0\,\sigma$. For the grid,
             different numbers of threads $n$ per particle were tested; the fastest time is reported
             on the left axis with the corresponding number of threads on the right axis.}
    \label{fig:ljN}
\end{figure}

We generated additional configurations for the Lennard-Jones fluid at $\rho = 0.6/\sigma^3$ with $N$ ranging
from 1,000 up to 128,000. We determined the optimal $n$ for the grid traversal and report only the fastest
total time. Figure~\ref{fig:ljN} shows the results on the Tesla V100 with double-precision particle data.
(Qualitatively similar results were obtained for most other benchmark configurations.) As expected, the
optimal $n$ for the grid increased as $N$ decreased. Nonetheless, the BVH consistently outperformed the
grid for all $N$, despite the effectively wider parallelism of the grid traversal. This indicates that the
BVH algorithm is well-suited even for multi-GPU simulations where strong scaling efficiency is needed.

The quantized BVHs provide superior performance to the grid, especially with double-precision particle data,
due to their decreased memory traffic and the higher throughput of single-precision arithmetic instructions. However,
compressing the node using integers reduces the tightness of the AABB bounds and may introduce
additional ``false-positive'' neighbors. Although this does not affect the correctness of the neighbor search
(no true neighbors are missed), it may increase the cost of other steps in a molecular simulation, e.g., force
calculation from a neighbor list, and it is desirable to keep the number of false-positive neighbors small.

We counted the average number of neighbors per particle found with the quantized
BVH, and compared it to the ``true'' number that lie within $r_{\rm c}$ as determined by the grid. We found for the Lennard-Jones
fluid that on average the quantized BVH identified 1.5 false-positive neighbors per particle at
$\rho=0.2/\sigma^3$ and 3.8 false-positive neighbors per particle at $\rho=0.8/\sigma^3$, corresponding to a
roughly 5\% increase in the total number of neighbors. This small number of false positives could be removed
by additional processing, or can simply be accepted provided that subsequent calculations using the neighbors
are not too computationally intensive.

Overall, the quantized BVH gave a speedup of 1.7x to 3.1x compared to the previous BVH algorithm \cite{Howard:2016wq}
and 1.1x to 2.3x compared to HOOMD-blue's atomic-operation grid algorithm on Tesla V100.
(Although the latter speedup is smaller than in Figs.~\ref{fig:ljdouble} and \ref{fig:ljsingle},
the quantized BVH has the additional benefits of consuming less memory and producing a deterministic neighbor
list.) To contextualize these values, we profiled the contribution of the neighbor search to the total simulation
time using HOOMD-blue's current algorithms with an optimized neighbor-list buffer width.
The neighbor search contribution ranged from as low as 20\% for the grid at $\rho = 0.2/\sigma^3$ to
nearly 50\% for the BVH at $\rho=0.8/\sigma^3$, meaning that the quantized BVH could speedup the total
simulation by as much as 1.5x on Tesla V100. In practice, though, the actual speedup may be smaller due
to additional overhead in the simulations and package-specific implementation details, which are not
present in our benchmarks.

\subsection{Weeks--Chandler--Andersen fluid}
Although the Lennard-Jones fluid is a standard simulation benchmark, many simulation models have shorter
interaction ranges than $r_{\rm c} = 3.0\,\sigma$. A common example is the Weeks--Chandler--Andersen (WCA) fluid \cite{WEEKS:1971uq},
which is a model for nearly-hard spheres obtained by truncating and shifting eq.~\ref{eq:lj} to zero at its minimum
($r_{\rm c} = 2^{1/6}\,\sigma$). Such short cutoffs are often used in fluids with very short-ranged or purely
repulsive interactions, but may prove challenging for the grid neighbor search for two reasons. First, if the
grid cell size is equal to $r_{\rm c}$, the cell occupancy becomes small as $r_{\rm c}$ decreases, and the overhead
of accessing a cell increases. Second, the number of cells in the grid may grow significantly as $r_{\rm c}$
decreases, limiting the volume of the simulation box that can be searched. The BVH suffers from neither of these issues
because it partitions the system based on objects rather than space.

\begin{figure}[!htbp]
    \centering
    \includegraphics{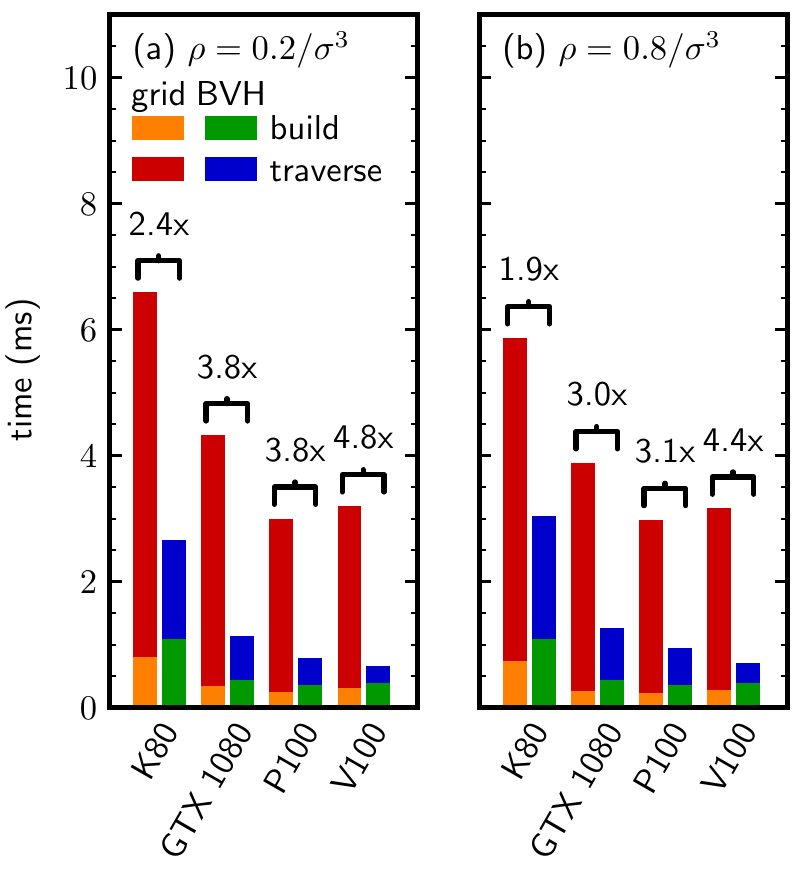}
    \caption{WCA fluid neighbor-search benchmark for $N = 128,000$ particles represented in double precision at
        (a) $\rho = 0.2/\sigma^3$ and (b) $\rho = 0.8/\sigma^3$ with $T = 1.5\,\varepsilon/k_{\rm B}$.
        The traversal counted the number of neighbors within a cutoff $r_{\rm c} = 2^{1/6}\,\sigma$.
        Speedups are the ratios of the total times rounded down to the nearest tenth.}
    \label{fig:wcadouble}
\end{figure}

We generated configurations of the WCA fluid with the same $N$, $T$, and $\rho$
as for the Lennard-Jones fluid. Figure~\ref{fig:wcadouble} shows the neighbor search times for the
double-precision particle data. The BVH outperformed the grid by a larger factor than for
the Lennard-Jones fluid (Fig.~\ref{fig:ljdouble}), consistent with our expectations that the grid should be less favorable with a shorter
cutoff radius. However, the speedup on GTX 1080 at the higher density (Fig.~\ref{fig:wcadouble}b) was smaller than
for the Lennard-Jones fluid (Fig.~\ref{fig:ljdouble}b) due to the lower cell occupancy, which resulted in fewer
direct distance evaluations per particle for the grid. Similar trends were obtained for the WCA fluid
as for the Lennard-Jones fluid using single-precision particle data instead of double-precision data,
and so these results are omitted here for brevity.

\subsection{Spinodal decomposition}
We finally tested for effects of inhomogeneity often encountered in self-assembly or phase coexistence simulations,
where it is common to have regions of higher or lower density within one simulation box. The BVH should adapt well to
such density variations because it partitions by objects rather than space, and so is expected to again perform
favorably compared to the grid. To generate configurations with density variations, we quenched the Lennard-Jones
fluid into the spinodal region of its phase diagram, where it spontaneously decomposes into regions of low density
(vapor) and high density (liquid). We first equilibrated a supercritical fluid of $N = 10^6$ particles at
$\rho = 0.2/\sigma^3$ and $T = 1.5\,\varepsilon/k_{\rm B}$. We then reduced the temperature below the critical point \cite{Potoff:1998,Panagiotopoulos:1994}
to $T = 0.8\,\varepsilon/k_{\rm B}$ and sampled configurations every $10^3\,\tau$ as the fluid underwent spinodal
decomposition. In addition to testing the impact of density variations, this benchmark also
assesses BVH performance for large $N$.

\begin{figure}[!htbp]
    \centering
    \includegraphics{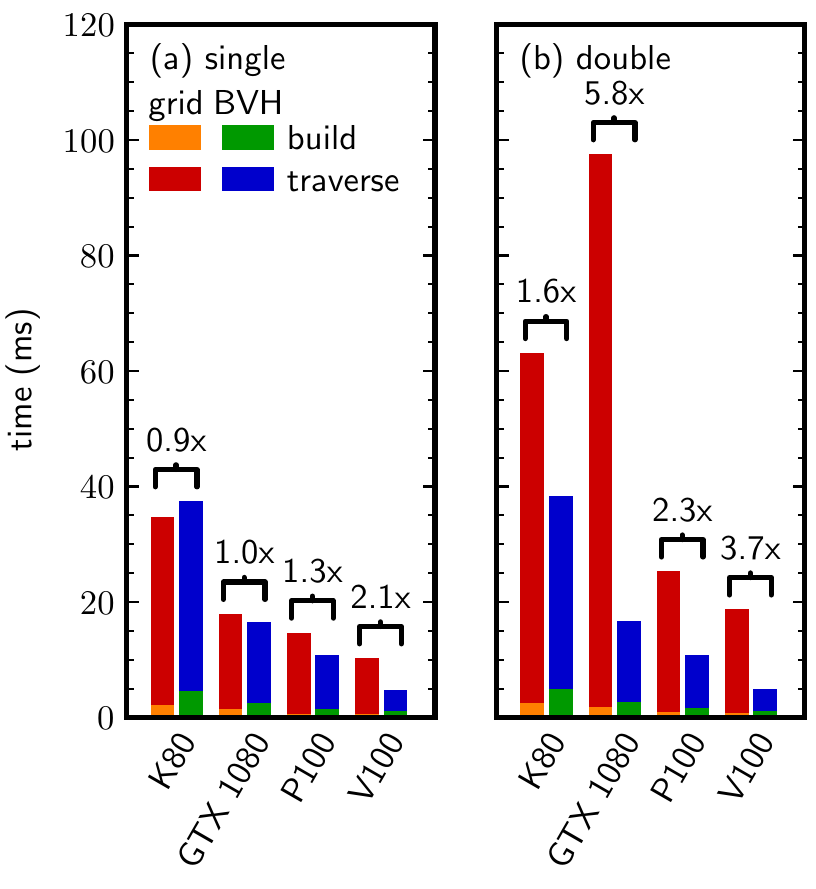}
    \caption{Spinodal decomposition benchmark for $N = 10^6$ particles represented in (a) single and (b) double precision. The
        fluid initially had $\rho = 0.2/\sigma^3$ and $T =1.5\,\varepsilon/k_{\rm B}$. The
        temperature was quenched to $T = 0.8\,\varepsilon/k_{\rm B}$, below the critical point of the
        Lennard-Jones fluid \cite{Potoff:1998,Panagiotopoulos:1994}, and the fluid decomposed into liquid and vapor regions.
        The traversal counted the number of neighbors within a cutoff $r_{\rm c} = 3.0\,\sigma$. Speedups
        are the ratios of the total times rounded down to the nearest tenth.}
    \label{fig:ljspinodal}
\end{figure}

Figure~\ref{fig:ljspinodal} reports the neighbor search times for both single-precision and double-precision
particle data. We again find that the BVH consistently outperformed the grid for nearly all GPUs and benchmark configurations.
The largest speedups were obtained for double-precision particle data
(Fig.~\ref{fig:ljspinodal}b), where the quantized BVHs are most advantageous due to the smaller node size compared
to the particle data and the higher single-precision arithmetic instruction throughput. However, good speedups were also obtained for the single-precision
particle data, again rendering the BVH the best choice for neighbor search in such systems.

\section{Conclusions} \label{sec:conc}
We developed an improved implementation of our previously proposed algorithm \cite{Howard:2016wq} for performing neighbor searches with BVHs
in molecular simulations using GPUs. A low-precision, quantized representation of the BVH
nodes significantly increased the BVH traversal speed. We showed with a suite of benchmarks that
the improved BVH neighbor search outperforms an equivalent grid (``cell list'') search on multiple generations
of NVIDIA GPUs, and is roughly two to four times faster on current GPUs. Given these benchmark results for
single-component fluids and our previous finding that BVHs outperform a single cell list for size-asymmetric pair
interactions in mixtures \cite{Howard:2016wq}, we recommend using the BVH instead of a single cell list to
generate neighbor lists in molecular simulations on GPUs.

Looking forward, the performance of the BVH neighbor search may be further improved by advances in software and
hardware. Given that the LBVH build time is still only a small fraction of the total neighbor-search time,
higher quality BVHs may improve the neighbor search performance. However, the longer times needed to build
such BVHs must be offset by significantly faster traversal. The strategies of ref.~\cite{Pall:2013} may be combined
with the BVH as a form of subtree collapse to reduce the number of traversals needed and to maximize parallelism.
A recently introduced NVIDIA GPU has specialized hardware for ray tracing, with dedicated units for BVH traversal
of rays against triangles in leafs \cite{nvidia:turing}. It may be possible to capitalize on this hardware to perform
neighbor searches in molecular simulations, possibly utilizing the free parts of the GPU to overlap other
calculations, if it can be programmed for more general purposes.
 
\section*{Acknowledgments}
We gratefully acknowledge the NVIDIA Corporation for providing access to the PSG Cluster to perform the
benchmarks on Tesla K80, P100, and V100. M.P.H. and T.M.T. acknowledge support from the
Welch Foundation (Grant No. F-1696). Financial support for this work (A.S., A.Z.P.) was partially provided
by the Princeton Center for Complex Materials, a U.S. National Science Foundation Materials Research Science
and Engineering Center (Award No. DMR-1420541).

\section*{Data availability}
The raw and processed data required to reproduce these findings is available from the corresponding
author (M.P.H.) upon request. The implementations of the algorithms benchmarked in this work are available
as open-source software \cite{neighbor}.

\section*{References}
\bibliography{ms}

\end{document}